\documentclass[aps,twocolumn,epsfig,rotate]{revtex4}
\usepackage{epsfig}
\usepackage{graphicx}
\usepackage{amsmath}

\newcommand{\beq}{\begin{eqnarray}}
\newcommand{\eeq}{\end{eqnarray}}
\begin{document}                

\title{Complete quantum control of the population transfer 
branching ratio between two degenerate
target states}
\author{Jiangbin Gong and Stuart A. Rice}
\affiliation{Department of Chemistry and The James Franck Institute,\\
The University of Chicago, Chicago, Illinois 60637}
\date{\today}

\begin{abstract}
A five-level four-pulse phase-sensitive extended stimulated Raman adiabatic  passage scheme is proposed to realize
complete control of the population transfer branching ratio between 
two degenerate target states. 
The control is achieved via a three-node null eigenstate that can be correlated with an arbitrary superposition of the target states.
Our results suggest that complete suppression of the yield of  one of two degenerate product states, and
therefore absolute
selectivity in photochemistry, is achievable and predictable,
even without studying the properties of the unwanted product state beforehand.
\end{abstract}
\maketitle

\section{Introduction}

Optical control of atomic and molecular processes has attracted great theoretical and experimental interest in recent years \cite{ricebook,brumerbook}.
The seminal study of Brumer and Shapiro \cite{brumer86}
showed that the relative phase
of two independent excitation pathways between the same
initial and target states
can be optically manipulated to generate interference control over the ratio of populations tranferred to degenerate product states.
However, this weak-field picture of the controlled quantum dynamics is not very useful in the case of strong fields, where
the number of interfering pathways that relate the initial state to the same
target state can be enormous.

The stimulated Raman adiabatic passage (STIRAP) \cite{bergmann} method for population transfer is a powerful strong field approach to
the control of atomic and molecular processes.
In its simplest three-level version, STIRAP generates coherent population transfer between two states of the same parity without populating
the intermediate state. The strong field eigenstates are conveniently represented as time-dependent superpositions of the three 
field-free states. 
The key element in the STIRAP process is the creation, via counter-intuitively ordered pump and Stokes pulses, 
of a null eigenstate
that has a node on the intermediate field-free state
and correlates with both the field-free initial and target states.
There have been a number of successful extensions of STIRAP to multi-level systems
\cite{shore91,tannor97,oreg92,coulston92,shore,vitanov,kobrakpra,kobrakjcp,shapiro,sola,unanyan0,unanyan,kis,kis2}.
In particular,
Kobrak and Rice \cite{kobrakpra} (KR) have proposed 
a five-level extended STIRAP scheme to achieve
product selectivity in photochemistry.  This scheme was later shown to be the essence of 
the important Chen-Shapiro-Brumer (CSB) strong field approach \cite{kobrakjcp,chenjcp,chenpra} to the control of 
photodissociation reactions (note that the CSB approach was experimentally realized in the Na$_{2}$ system \cite{chenpra}).
Since then, there has been
a renewed interest in studies of controlled molecular dynamics from a STIRAP perspective
\cite{rice01,kurkal01a,kurkal01b,kurkal02,rice02,
mustafa,mustafa2,gong2,gong3,gong4,ohta,geva,vrabel}.

The central interest in the KR scheme is to control 
the population transfer branching ratio between two degenerate target states. This issue is of great importance to selective photochemistry
because photodissociation reactions typically form degenerate product states,
with each product state correlated with a different potential energy surface. Unlike the degeneracy associated with, for example,
different magnetic sublevels where the exact degeneracy can be lifted by Zeeman splitting \cite{shore}
and certain sublevels may be addressed individually by taking advantage of the selection rules associated with 
polarized laser fields \cite{unanyan0,unanyan},
the exact degeneracy between different chemical product states in the continuum cannot be lifted. Then,
if one of the degenerate product states is
coupled with a particular state by a laser field with an arbitrary polarization, so is the other product state.  Indeed,
neither the KR scheme nor the CSB strong field approach offers a means for completely
suppressing population transfer to an arbitrarily chosen product channel.
A simpler and potentially very useful approach to control of population transfer that 
takes advantage of a large lifetime difference between two degenerate product states has been proposed \cite{gong3}, but
the control is incomplete when both target states rapidly decay. 

Recently, we advocated a measurement-based approach that may be used to completely shut off the yield of one of two degenerate product states \cite{gong4}. In particular,
we have shown that a strong  measurement of the population of the branch state in the KR
scheme modifies the spectrum of the system such that the nonadiabatic coupling and the quantum interference
between two particular adiabatic states can be manipulated. This approach to population transfer control
overlaps with the study of the quantum Zeno and anti-Zeno effects.
However, this approach does not totally resolve the issue of achieving complete control of the population transfer branching ratio.
In fact, the typical yield of the desired product state is not 100\% and analytical conditions for
 the complete suppression of the yield of one product state
remain to be established.

In this paper, we propose a five-level  four-pulse extended STIRAP scheme that offers
complete control of the population transfer branching ratio between two degenerate product states.  The results strongly suggest that
complete selectivity of product formation is in principle achievable.
As shown below,  in addition to introducing one more laser pulse than in the KR model and in our measurement-based approach,
the price that is paid for complete control 
is that the relative phases between the four laser pulses must be well maintained. 
This requirement is experimentally demanding,
but with the evolution of laser technology we expect it to be met.
Even if the relative phases between the four laser pulses cannot now be controlled, we expect that 
a number of attractive properties
inherent to our five-level  four-pulse scheme will motivate new experiments.
For example, we demonstrate that
one does not need to analyze the properties of one of the two degenerate product states in order to
predict the complete suppression of its yield. We also demonstrate that an arbitrary superposition of the two degenerate
target states can  be created.  Finally,
we show that complete suppression of the yield of one of two degenerate product states
can be realized  when the product states have  lifetimes that are short
compared with the duration of the laser pulses.

The achievement of complete controllability in multi-level systems may be rigorously formulated
as a mathematical problem. Some existence proofs 
that establish sufficient conditions for complete control 
have been discovered \cite{huang1,rama,tersigni}.  
The Huang-Tarn-Clark theorem
\cite{huang1} is believed to the strongest result, but it
applies only to systems with discrete and nondegenerate states.
Ramakrishna
{\it et al.} showed that for a quantum system with a Hilbert space of dimension $L$,
the necessary and also sufficient condition for complete controllability is that the
field-free Hamiltonian and the interaction Hamiltonian induced by a
control field
generate a Lie algebra of dimension $L^{2}$ \cite{rama}. This criterion is applicable
to degenerate multi-level systems,
but the required
computations to generate the commutators of the Lie algebra structure can be demanding
for large $L$,  vary drastically from system to system,
and provide no hint as to how a control field can be constructed.
Even if the existence of 
complete controllability
in a multi-level system that involves degenerate states is established using a Lie algebra analysis,
that result is unlikely to provide hints for the following:
(i) how to design a conceptually simple control
scheme to realize the complete control; (ii) how a physically valid approximation changes the picture of controllability,
(iii) how the decay of product states affects the controllability,
and (iv) how the controllability is influenced by constraints on the system.
Consider, for example, 
the generation of a complete
population transfer pathway in a five-level system that never
populates more than two levels. This control scheme
can be formulated as a mathematical question that
is very different from 
the original question of how to achieve complete controllability in the same system.
Indeed, 
as suggested by an early study by Shapiro and Brumer \cite{shapirojcp},
such a population transfer pathway is very unlikely to exist (but not necessarily impossible) because the
dimension of the Hilbert subspace that is disallowed by 
the constraint is too large relative to that of
the rest of the Hilbert space.
Nevertheless, as we show below in a straightforward manner, 
a complete population transfer pathway in a five-level system
that does not populate three levels does exist, indicating that the achievement of complete controllability  may be
possible with reasonable approximations (such as the adiabatic approximation and the rotating-wave approximation)
even though the underlying mathematical question seems difficult to resolve or suggests the opposite.
Hence, our five-level four-pulse control scheme is also of great theoretical interest
in the context of
the formal question of complete controllability without or with strong constraints.

This paper is organized as follows. In Sec. II we describe, in connection with
the original KR five-level model, our five-level four-pulse phase-sensitive extended STIRAP scheme for the complete control
of the population transfer branching ratio between two degenerate product states.
In Sec. III we show that our control scheme can also be used for the creation of arbitrary  superpositions of two degenerate target states.
Some simple computational examples that support our theoretical analysis are presented in Sec. IV.
Our results are discussed and set in context in Sec. V.

\section{A five-level four-pulse extended STIRAP scheme}
\subsection{The model system}
\begin{figure}[ht]
\ \vspace{-0.5cm} \begin{center}
\epsfig{file=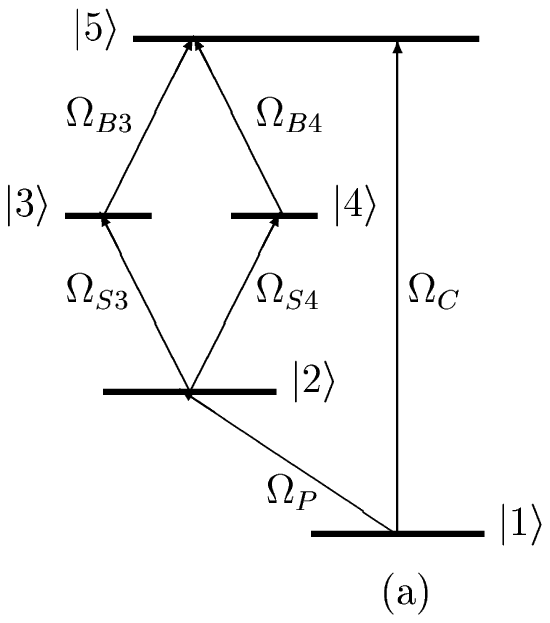,width=6.5cm}
\epsfig{file=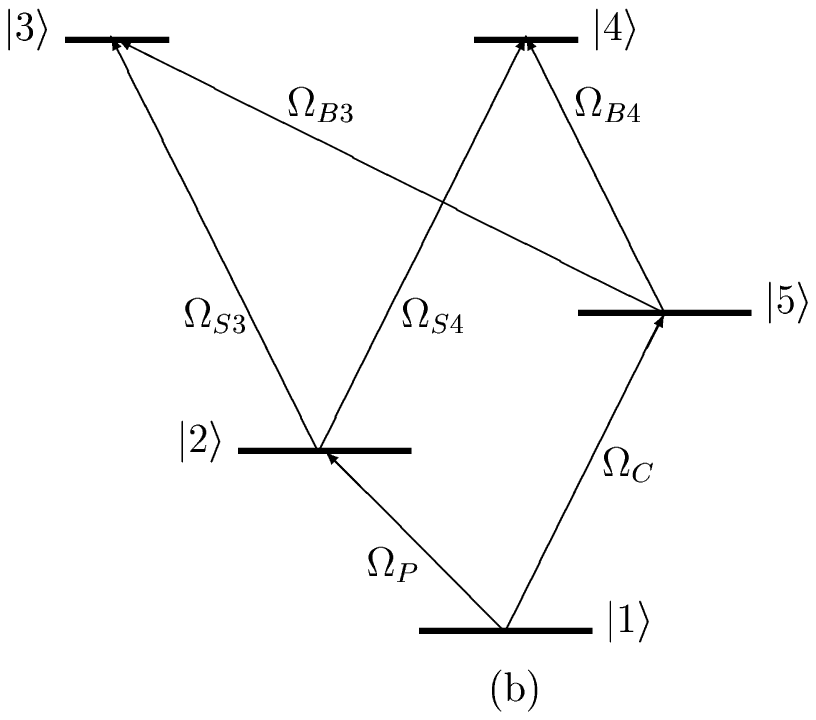,width=6.5cm}
\end{center}
\caption{A schematic diagram of the five-level four-pulse phase-sensitive extended STIRAP method for
the complete control of the population transfer branching ratio between two degenerate product states $|3\rangle$
and $|4\rangle$.  The initial state is $|1\rangle$. $\Omega_{P}$, $\Omega_{C}$, $\Omega_{S3}$ and  $\Omega_{S4}$,
$\Omega_{B3}$ and  $\Omega_{B4}$ represent the Rabi frequencies associated with the pump pulse, the control pulse, the
Stokes pulse, and the branching pulse, respectively.  Under certain conditions 100\% population can be transferred from
state $|1\rangle$ to state $|4\rangle$ without ever populating states $|2\rangle$, $|3\rangle$, or $|5\rangle$.
Cases (a) and (b) represent two possible level configurations.}
\label{fig1}
\end{figure}

A schematic diagram
of our five-level four-pulse scheme is shown
in Fig. \ref{fig1}. 
The initial state $|1\rangle$ is coupled with the intermediate state $|2\rangle$ by a pump pulse,
state $|2\rangle$ is coupled with two degenerate product states $|3\rangle$ and $|4\rangle$ by a Stokes pulse.
As in our previous work \cite{kobrakpra,kobrakjcp}, we have adopted the common  (but less rigorous)
treatment wherein the product states $|3\rangle$ and $|4\rangle$, if embedded in the continuum, are described
as discrete levels with complex energies whose imaginary parts give the decay rate constants.
Unless specified otherwise below, we first assume that the lifetimes of all the five levels are much longer than the
duration of the laser pulses.
This assumption is not essential for the success of our control scheme,
but is convenient in the description of the physical picture of
the control
mechanism.
Due to the degeneracy of states $|3\rangle$ and $|4\rangle$,  alteration of their population transfer branching ratio (denoted by $B$) 
cannot be achieved unless other laser-induced couplings are introduced.
This observation motivated  Kobrak and Rice to introduce 
a ``branch state" $|5\rangle$ that is also coupled with states $|3\rangle$ and $|4\rangle$ by a third ``branching
pulse''.   
 Here we consider also a fourth laser pulse, called  the ``control pulse'',
that further couples the branch state with the initial state.
From Fig. \ref{fig1} it is seen that  the control
pulse introduces  an evident  symmetry to  the system: state $|5\rangle$ can be thought of as a second intermediate state,
and the control (branching) pulse plays a similar role to that of the pump (Stokes) pulse.
As shown below, this extension from the KR
model to our five-level four-pulse scheme leads to a new class
of STIRAP-like processes that can be advantageously manipulated by laser phases.

We assume that all the laser fields are on resonance with the respective transitions and that the pulse shapes are Gaussian.
The electric fields of the pulses are given by
\beq
E_{X}(t)=\tilde{E}_{X}f_{X}(t)\cos(\omega_{X}t+\phi_{X}),
\label{e1}
\eeq
with the envelope function 
\beq
f_{X}(t)=\exp\left[-(t-t_{X})^{2}/T_{X}^{2}\right].
\label{e2}
\eeq
In Eqs. (\ref{e1}) and (\ref{e2}), $\omega_{X}$ is the frequency of the laser pulse,
$\phi_{X}$ is the associated phase, $T_{X}$ is the pulse width, $t_{X}$
represents the timing of the peak electric field amplitude $\tilde{E}_{X}$,
and $X=P, S, B, C$ for the pump, Stokes, branching, and control pulses, respectively.
Below we confine ourselves to level configuration (b) in Fig. \ref{fig1}; the other configuration can be analyzed in a similar
fashion.
Using the rotating wave approximation (for discussions on the validity of this approximation, see Ref. \cite{kurkal01b})
and the interaction representation,
the system Hamiltonian is given by
\begin{equation}
H=\left[
\begin{array}{ccccc}
0 & \Omega_{P} & 0  & 0 &  \Omega_{C} \\
\Omega_{P}^{*} & 0 & \Omega_{S3} &  \Omega_{S4} & 0 \\
0 & \Omega_{S3}^{*} & 0 & 0 & \Omega_{B3} \\
0 &  \Omega_{S4}^{*}& 0 &  0 & \Omega_{B4} \\
\Omega_{C}^{*} & 0 & \Omega_{B3}^{*} & \Omega_{B4}^{*} & 0\\
\end{array}
\right].
\end{equation}
Here the six Rabi frequencies (see Fig. \ref{fig1}b) are 
\beq
\Omega_{P}&=&|\mu_{12}|(\tilde{E}_{P}/2)f_{P}(t)\exp[i(\phi_{P}+\alpha_{12})],\\
\Omega_{S3}&=&|\mu_{23}|(\tilde{E}_{S}/2)f_{S}(t)\exp[i(\phi_{S}+\alpha_{23})], \label{u1}\\
\Omega_{S4}&=&|\mu_{24}|(\tilde{E}_{S}/2)f_{S}(t)\exp[i(\phi_{S}+\alpha_{24})],\\
\Omega_{B3}&=&|\mu_{35}|(\tilde{E}_{B}/2)f_{B}(t)\exp[-i(\phi_{B}-\alpha_{35})],\\
\Omega_{B4}&=&|\mu_{45}|(\tilde{E}_{B}/2)f_{B}(t)\exp[-i(\phi_{B}-\alpha_{45})], \label{u4}\\
\Omega_{C}&=&|\mu_{15}|(\tilde{E}_{C}/2)f_{C}(t)\exp[i(\phi_{C}+\alpha_{15})],
\eeq
where $|\mu_{kj}|$ and $\alpha_{kj}$ represent the magnitude and the phase
of the transition dipole moment $\mu_{kj}$ between states $|k\rangle$ and $|j\rangle$.

\subsection{The null eigenvector}
It is quite unexpected and fortunate
that  the eigenvalues and eigenvectors of $H$, denoted by
$\lambda_{k}$ and $|\lambda_{k}\rangle$ ($k=1-5$) below,
can be obtained analytically.
In particular, we find that $H$ has one null eigenvalue (denoted  $\lambda_{1}$), with
the associated eigenvector given by
\beq
|\lambda_{1}\rangle=\frac{1}{N_{1}}\left[
\begin{array}{c}
\Omega_{S4}\Omega_{B3}^{*}-\Omega_{S3}\Omega_{B4}^{*}\\
0 \\
\Omega_{P}^{*}\Omega_{B4}^{*}-\Omega_{S4}\Omega_{C}^{*}\\
\Omega_{S3}\Omega_{C}^{*}-\Omega_{P}^{*}\Omega_{B3}^{*}\\
0\\ \end{array}
\right],
\label{eq1}
\eeq
where $N_{1}$ is the normalization factor. Except for one accidental case discussed below,
the other eigenvalues are found to be nonzero, with $\lambda_{k}^{2}$ ($k=2-5$)
satisfying a simple ``quadratic" equation, namely
\beq
0& = & \left(\lambda_{k}^{2}\right)^{2}-\lambda_{k}^{2}\Omega^{2}+
\left|\Omega_{S3}\Omega_{B4}^{*}-\Omega_{S4}\Omega_{B3}^{*}\right|^{2} \nonumber \\
&& + |\Omega_{P}^{*}\Omega_{B4}^{*}-\Omega_{C}^{*}\Omega_{S4}|^{2}
+ |\Omega_{P}^{*}\Omega_{B3}^{*}-\Omega_{C}^{*}\Omega_{S3}|^{2},
\label{l25}
\eeq
where $\Omega^{2}$ is the sum of the absolute squares of all the six Rabi frequencies. 
%
Using Eq. (\ref{l25}) the analytical forms for $|\lambda_{k}\rangle$ ($k=2-5$) can also be
obtained. However,
since states $|\lambda_{k}\rangle$ ($k=2-5$) are in general not degenerate with $|\lambda_{1}\rangle$, the nonadiabatic coupling
between $|\lambda_{k}\rangle$ ($k=2-5$) and $|\lambda_{1}\rangle$ can be made negligible by use of sufficiently strong fields.
With the weak nonadiabaticity assumption,
only the null eigenstate is relevant to theoretical consideration of the dynamics of population transfer.

Consider first two special cases in which our scheme reduces to the original KR model where only
three pulses are used.
In the first case $\Omega_{C}=0$, $\Omega_{P}\ne 0$, and state $|5\rangle$ is the  branch state. Then one has
\beq
|\lambda_{1}\rangle\rightarrow |0\rangle_{KR}\equiv \frac{1}{N_{1}}\left[\begin{array}{c}
\Omega_{S4}\Omega_{B3}^{*}-\Omega_{S3}\Omega_{B4}^{*}\\
0 \\
\Omega_{P}^{*}\Omega_{B4}^{*}\\
-\Omega_{P}^{*}\Omega_{B3}^{*}\\
0\\ \end{array}\right].
\label{01kr}
\eeq
The associated branching ratio is given by
$B=|\Omega_{B4}|^{2}/|\Omega_{B3}|^{2}=|\mu_{45}|^{2}/|\mu_{35}|^{2}$.
Similarly, 
if $\Omega_{P}=0$,  $\Omega_{C} \ne0$, then state $|2\rangle$ 
plays the role of the branch state and
\beq
|\lambda_{1}\rangle\rightarrow |0\rangle_{KR}'\equiv \frac{1}{N_{1}}
\left[\begin{array}{c} \Omega_{S4}\Omega_{B3}^{*}-\Omega_{S3}\Omega_{B4}^{*}\\
0\\
-\Omega_{C}^{*}\Omega_{S4}\\
\Omega_{C}^{*}\Omega_{S3}\\
0\\ \end{array} \right].
\label{02kr}
\eeq
The associated value of 
$B$ is then given by $|\Omega_{S4}|^{2}/|\Omega_{S3}|^{2}=|\mu_{24}|^{2}/|\mu_{23}|^{2}$.
Clearly, the value of $B$ in either case 
is determined entirely by the properties of state $|5\rangle$ or state $|2\rangle$,
is a constant during the population transfer, and does not carry any information about the laser phases.
Further, for each desired value of $B$ one has to seek a different branch state and then adjust the laser frequencies, or, alternatively,
detune the branching pulse from the branch state in an effort to generate the desired value of $B$.
Unless one of the transition dipole moments  $\mu_{35}$, $\mu_{45}$, $\mu_{23}$ and  $\mu_{24}$ happens to be zero, 
complete suppression of the yield of one product state
or varying the value of $B$ from $0$ to $\infty$ is generally out of the question.
In this sense, control of the population transfer branching ratio is incomplete.

Consider now the null eigenvector $|\lambda_{1}\rangle$ with $\Omega_{P}\ne 0$ and $\Omega_{C}\ne 0$.
First, if the pump pulse follows the
Stokes pulse and the control pulse follows the branching pulse, then $|\lambda_{1}\rangle$ will
first correlate with the initial state
and then correlate with the product states as time evolves. Second, 
the two nodes of $|\lambda_{1}\rangle$ on
states $|2\rangle$  and $|5\rangle$ are independent of the phases and amplitudes of the control pulse and the pump pulse.
Hence, while $|\lambda_{1}\rangle$ remains the key eigenstate for adiabatic population
transfer whilst maintaining zero population of the intermediate state and the branch state, 
it possesses additional experimentally controllable parameters
that are absent in $|0\rangle_{KR}$ and  $|0\rangle_{KR}'$.
Third, it is seen that 
the value of $B$ associated with $|\lambda_{1}\rangle$ can be time-dependent during the population transfer
and is sensitive to the relative phases of the laser pulses. 
This phase sensitivity is intriguing and makes it clear that
our five-level four-pulse phase-dependent extended STIRAP scheme
is a marriage between phase-insensitive STIRAP and
coherent phase control. Indeed, comparing Eq. (\ref{eq1}) with Eqs. (\ref{01kr}) and (\ref{02kr}),
one sees that  the null eigenvector $|\lambda_{1}\rangle$ in Eq. (\ref{eq1})  may be understood as a coherent superposition of
the two null eigenvectors $|0\rangle_{KR}$ and $|0\rangle_{KR}'$ associated with the KR model. That is, in essence, we are able to
utilize two KR schemes simultaneously by adding one more laser pulse to the system.  
Our five-level four-pulse approach is also related to and, as seen below, more ambitious than,
a recent proposal by Karpati and Kis to achieve selective adiabatic population transfer to degenerate magnetic sublevels  \cite{kis}.

\subsection{Complete yield suppression of one product state}
The key element in our approach to controlling population  transfer
is motivated by the observation that the laser pulses can be designed such that
a third node can be introduced to the null eigenvector.  From Eq. (\ref{eq1})
it is seen that if the control fields can ensure that
\beq 
\Omega_{P}^{*}\Omega_{B4}^{*}-\Omega_{S4}\Omega_{C}^{*}=0,
\label{con1}
\eeq
then
\beq
|\lambda_{1}\rangle=\frac{(\Omega_{S4}\Omega_{B3}^{*}-\Omega_{S3}\Omega_{B4}^{*})}{N_{1}\Omega_{B4}^{*}}\left[
\begin{array}{c}
\Omega_{B4}^{*}\\
0 \\
0\\
-\Omega_{C}^{*}\\
0\\ \end{array}
\right].
\label{eq2}
\eeq
That is,  $|\lambda_{1}\rangle$ will have an additional node on state $|3\rangle$. A similar condition {\it i.e.,} 
$\Omega_{S3}\Omega_{C}^{*}-\Omega_{P}^{*}\Omega_{B3}^{*}=0$
will induce a node on state $|4\rangle$.
With this third node of $|\lambda_{1}\rangle$ transfer of population to
one of the degenerate product states
can be completely shut off.  Significantly, since the third node on state  $|3\rangle$ (if obtained) implies
that the time-evolving adiabatic state $|\lambda_{1}\rangle$
will have zero overlap with state $|3\rangle$ at all times,
it becomes possible that state $|3\rangle$ 
is never populated during the population transfer from the initial state to the target state $|4\rangle$. 

Condition (\ref{con1}) can be met at all times by controlling the phases and the envelope functions of the four laser pulses.
As the Rabi frequencies are complex in general and time-dependent,
condition (\ref{con1}) requires matches
of both the laser phases $\phi_{X}$,
and their envelope functions $f_{X}(t)$. Specifically, 
condition (\ref{con1}) holds if
\beq
\phi_{B}+\phi_{C}-\phi_{P}-\phi_{S}= \alpha_{45}+\alpha_{24}+\alpha_{12}-\alpha_{15}, \label{c0} \\
\tilde{E}_{P}\tilde{E}_{B}|\mu_{12}||\mu_{45}|\exp\left[-\frac{(t_{P}-t_{B})^{2}}{T_{P}^{2}+T_{B}^{2}}\right] \nonumber \\
 =\tilde{E}_{S}\tilde{E}_{C}|\mu_{24}||\mu_{15}|\exp\left[-\frac{(t_{S}-t_{C})^{2}}{T_{S}^{2}+T_{C}^{2}}\right], \label{c3} \\
\frac{T_{P}^{2}+T_{B}^{2}}{T_{P}^{2}T_{B}^{2}}=\frac{T_{S}^{2}+T_{C}^{2}}{T_{S}^{2}T_{C}^{2}},\label{c1} \\
\frac{T_{B}^{2}t_{P}+T_{P}^{2}t_{B}}{T_{P}^{2}+T_{B}^{2}}=\frac{T_{S}^{2}t_{C}+T_{C}^{2}t_{S}}{T_{S}^{2}+T_{C}^{2}}. \label{c2} 
\eeq
Clearly, with the intensity and phases of all other laser pulses fixed, Eqs. (\ref{c0}) and (\ref{c3}) can be
fulfilled by scanning the  phase and intensity of the control pulse. 
Equations (\ref{c1}) and (\ref{c2}) are independent of the molecular properties and can also be easily satisfied. 
We now consider 
two sample solutions to Eqs. (\ref{c1}) and (\ref{c2}) that will be used in the next section.
First, suppose that all the pulse widths are the same
{\it i.e.,} $T_{P}=T_{S}=T_{B}=T_{C}=T$. Then Eq. (\ref{c1}) is true; and if
$t_{S}=0$, $t_{P}=T$, $t_{B}=0$, then from Eq. (\ref{c2})  one obtains $t_{C}=T$.  Second, consider
another example where the branching pulse sits in the middle of
the pump and the Stokes pulses {\it i.e.,}  $t_{S}=0$, $t_{P}=T$, $t_{B}=0.5T$, and $T_{B}=\sqrt{2}T$, $T_{P}=T_{S}=T$.
Then one obtains $T_{C}=\sqrt{2}T$ from  Eq. (\ref{c1})
and $t_{C}=2.5T$ from  Eq. (\ref{c2}).
Note that these sample solutions
have ensured that the pulse order is counter-intuitive 
so that $|\lambda_{1}\rangle$ initially has unity overlap with state $|1\rangle$. 
Detailed conditions for completely suppressing population transfer to state $|4\rangle$  can also be found 
by slightly modifying Eqs. (\ref{c0}) and (\ref{c3}).

There is only one restriction imposed on the transition dipole moments in order to successfully apply
the five-level four-pulse approach outlined above.  In obtaining Eq. (\ref{eq1}) we have assumed that  
$\Omega_{S4}\Omega_{B3}^{*}\ne\Omega_{S3}\Omega_{B4}^{*}$. If
$\Omega_{S4}\Omega_{B3}^{*}=\Omega_{S3}\Omega_{B4}^{*}$ then Eq. (\ref{eq1}) suggests that
$|\lambda_{1}\rangle$ has a node
on state $|1\rangle$. 
In this case $|\lambda_{1}\rangle$ cannot be used for adiabatic population transfer
since it cannot correlate with the initial state. Indeed,
if we now further impose condition (\ref{con1}), 
then one obtains a third relation
$\Omega_{P}^{*}\Omega_{B3}^{*}=\Omega_{S3}\Omega_{C}^{*}$.  According to Eq. (\ref{l25}), this accidental case then results in
two additional null eigenstates that will
spoil the branching ratio control.  
To avoid $\Omega_{S4}\Omega_{B3}^{*}=\Omega_{S3}\Omega_{B4}^{*}$, one obtains from Eqs. (\ref{u1})-(\ref{u4}) 
the restriction
\beq
\frac{|\mu_{23}|}{|\mu_{24}|}\exp[i(\alpha_{23}-\alpha_{24})]\ne \frac{|\mu_{35}|}{|\mu_{45}|}\exp[-i(\alpha_{35}-\alpha_{45})].
\label{in1}
\eeq
Note that the above inequality also guarantees that if $|\lambda_{1}\rangle$ has a node on state $|3\rangle$ (or $|4\rangle$)
it will not have a node on state $|4\rangle$ (or $|3\rangle)$.

It should be stressed that given Eqs. (\ref{c0})-(\ref{in1})
none of the three states $|2\rangle$, $|3\rangle$, and $|5\rangle$ will, in the adiabatic limit,
be populated during the population transfer, and yet there is
100\% population
transfer from the initial state to state $|4\rangle$. 
Consider now more realistic cases where the lifetimes of the product states are short compared
with the duration of the laser pulses.   We assume that 
the control fields are sufficiently strong that
the associated peak Rabi frequencies are large compared with
the decay rate constants of the product states. Then
the lifetime effects can be described by
a first-order perturbative treatment \cite{kobrakjcp,gong4}. In the adiabatic limit this gives
\beq
P_{3}
=2\Gamma_{3}\int_{t_{i}}^{t_{f}}|\langle \lambda_{1}|3\rangle |^{2}\ dt,
\label{p3}
\eeq
where $\Gamma_{3}$ is the decay rate constant of state $|3\rangle$ and $P_{3}$ is the associated yield.
Equation (\ref{p3}) indicates that if the adiabatic state $|\lambda_{1}\rangle$ populates state $|3\rangle$ at some intermediate time,
then $P_{3}$ is necessarily nonzero.  That is, if state $|3\rangle$ decays, then  the mere condition
that the adiabatic state $|\lambda_{1}\rangle$ does not overlap with state $|3\rangle$ at later times does not suffice to
achieve complete suppression of $P_{3}$. 
However, if $|\lambda_{1}\rangle$ has a node on state $|3\rangle$, then  $\langle\lambda_{1}|3\rangle=0$ at all times
and 
Eq. (\ref{p3}) gives
$P_{3}=0$.
Clearly, then,  condition (\ref{con1}) for the complete suppression of $P_{3}$ is also valid
when the product states decay, and their decay rate constants are relatively small compared with the
 various peak Rabi frequencies (for example, if the lifetime of the target states is of the order of 100 femtoseconds,
then the required peak laser intensity should be of the order of $10^{11}$ W/cm$^{2}$).
That is,
since the unwanted
state $|3\rangle$ can be never populated, then the complete suppression of its yield  can also be achieved 
even when the lifetimes of the product states are short compared
with the duration of the laser pulses.  

The level configuration shown in Fig. \ref{fig1}b resembles that in the well-known 
``two-photon $+$ two-photon''
coherent phase control scenario \cite{brumerbook,chenjcp2}. However, the associated control mechanisms are drastically different.
In accord with the weak-field physical picture associated with the `two-photon + two-photon'' approach,
one intuitively expects that if $|\Omega_{P}|\cdot|\Omega_{S3}|=|\Omega_{C}|\cdot|\Omega_{B3}|$
then the interference between
the two excitation pathways $|1\rangle\rightarrow |2\rangle \rightarrow |3\rangle$ and  $|1\rangle\rightarrow |5\rangle \rightarrow |3\rangle$
is most effective and the phase control will be optimal.
This intuition is irrelevant and useless here, since condition (\ref{con1})
is a remarkable strong-field result and is extremely counter-intuitive. 
For example, condition (\ref{con1}), which is for the complete suppression of the yield of state 
$|3\rangle$, does not
specifically contain any information about the transition dipole moments that are related to state $|3\rangle$. 
In this sense the control is {\it universal}: as long as conditions (\ref{con1}) and (\ref{in1}) are met  and the dynamics
is  adiabatic, 
a state that is degenerate with, and orthogonal to, state $|4\rangle$ is guaranteed to have a zero yield,
even if  detailed information about the properties of this unwanted state is unavailable.
Another important difference between the ``two-photon $+$ two-photon'' scenario and
our five-level four-pulse scheme is that the former can be simplified in two special cases. Specifically,
if the energy difference between states $|5\rangle$ and $|1\rangle$ is identical
with that between states $|4\rangle$ and $|2\rangle$, or if state $|5\rangle$ is degenerate with state $|2\rangle$,
then
only two laser pulses are needed to realize the weak-field ``two-photon $+$ two-photon'' 
coherent control.
However, using only two laser pulses in these two special cases is not feasible
here due to the conditions required for adiabatic passage [see Eqs. (\ref{eq2}), (\ref{c0})-(\ref{c2})].  

\section{Creation of arbitrary superposition of two degenerate target states}
Of course,  states $|3\rangle$ and  $|4\rangle$
are just two candidate basis states of a two-dimensional degenerate subspace.
Consider now two new and arbitrary orthogonal basis states of this subspace,
\beq
|3'\rangle&=&\sin(\theta)|3\rangle + \exp(i\beta)\cos(\theta)|4\rangle, \\
|4'\rangle&=&\cos(\theta)|3\rangle - \exp(i\beta)\sin(\theta)|4\rangle, \label{s4}
\eeq
where $\theta$ and $\beta$ are two independent parameters characterizing the superposition states 
$|3'\rangle$ and  $|4'\rangle$.  
The transition dipole moments related to states $|3'\rangle$ and $|4'\rangle$, denoted by
$\mu_{23}'$, $\mu_{24}'$, $\mu_{35}'$, and $\mu_{45}'$,   can be obtained from
linear combinations of $\mu_{23}$, $\mu_{24}$, $\mu_{35}$, and $\mu_{45}$. Specifically,
\beq
\mu_{23}'&=&\sin(\theta)\exp(i\alpha_{23})|\mu_{23}| \nonumber \\
&& +\exp[i(\beta+\alpha_{24})]\cos(\theta)|\mu_{24}|, \\
\mu_{24}'&=&\cos(\theta)\exp(i\alpha_{23})|\mu_{23}|\nonumber \\
&& -\exp[i(\beta+\alpha_{24})]\sin(\theta)|\mu_{24}|, \\
\mu_{35}'&=&\sin(\theta)\exp(i\alpha_{35})|\mu_{35}|\nonumber \\
&& +\exp[-i(\beta-\alpha_{45})]\cos(\theta)|\mu_{45}|, \\
\mu_{45}'&=&\cos(\theta)\exp(i\alpha_{35})|\mu_{35}|\nonumber \\
&& -\exp[-i(\beta-\alpha_{45})]\sin(\theta)|\mu_{45}|.
\eeq
Now if we
apply the formalism outlined in the previous section to 
states $|3'\rangle$ and $|4'\rangle$ instead of states $|3\rangle$ and  $|4\rangle$, 
we find that the yield of state $|3'\rangle$ can be completely suppressed, resulting in $100\%$ population transfer
to state $|4'\rangle$. 
Not surprisingly, the specific conditions for  the complete suppression of the yield of state
$|3'\rangle$ are the same as Eqs. (\ref{c0})-(\ref{c2}), except that
$|\mu_{45}|\rightarrow |\mu_{45}'|$, $|\mu_{24}|\rightarrow |\mu_{24}'|$, $\alpha_{24}\rightarrow \alpha_{24}'$,
and $\alpha_{45}\rightarrow \alpha_{45}'$, where
\beq
|\mu_{24}'|=\sqrt{\cos^{2}(\theta)|\mu_{23}|^{2}+\sin^{2}(\theta)|\mu_{24}|^{2}},\\
|\mu_{45}'|=\sqrt{\cos^{2}(\theta)|\mu_{35}|^{2}+\sin^{2}(\theta)|\mu_{45}|^{2}},\\
\tan(\alpha_{24}')= \nonumber \\
 \frac{\cos(\theta)\sin(\alpha_{23})|\mu_{23}|-\sin(\theta)\sin(\beta+\alpha_{24})|\mu_{24}|}{\cos(\theta)\cos(\alpha_{23})|\mu_{23}|-
\sin(\theta)\cos(\beta+\alpha_{24})
|\mu_{24}|}, \\
\tan(\alpha_{45}')= \nonumber \\
 \frac{\cos(\theta)\sin(\alpha_{35})|\mu_{35}|+\sin(\theta)\sin(\beta-\alpha_{45})|\mu_{45}|}{
\cos(\theta)\cos(\alpha_{35})|\mu_{35}|-\sin(\theta)\cos(\beta-\alpha_{45})|\mu_{45}|}.
\eeq
Thus, an arbitrary superposition of two degenerate target states, such as
state $|4'\rangle$ in Eq. (\ref{s4}), can be created with our five-level four-pulse scheme if the lifetimes of states $|3\rangle$ and  $|4\rangle$  are negligible compared with
the duration of the laser pulses. Evidently, this also indicates that an arbitrary population transfer
branching ratio between two degenerate product states can be attained.


Our approach to the creation of an arbitrary superposition of two degenerate target states 
has two important advantages.  First, the unwanted
superposition state $|3'\rangle$, as discussed above,
can never be populated because the control is realized via the time-evolving adiabatic state that has a node on state $|3'\rangle$.
Second, it is
assumed that any single laser field cannot individually address the degenerate target states.  
In contrast, previous proposals \cite{unanyan0,unanyan,kis,kis2} for the creation of superposition states of magnetic sublevels
have at most only one of the two features.

The ability to create an arbitrary superposition of two degenerate molecular states is of considerable theoretical and experimental interest.
Suppose
state $|3\rangle$ is associated with one potential energy surface and state $|4\rangle$ is associated with
another potential energy surface, {\it i.e.,} $|3\rangle=|{\bf n}_{3}\rangle\otimes |e_{3}\rangle$
and $|4\rangle=|{\bf n}_{4}\rangle\otimes |e_{4}\rangle$, where $|{\bf n}_{3}\rangle$ and $|{\bf n}_{4}\rangle$
describe the rovibrational motion of states $|3\rangle$ and $|4\rangle$,
and $|e_{3}\rangle$ and $|e_{4}\rangle$ denote the associated electronic states.
One then obtains that the superposition state $|4'\rangle$ is an entangled state:
\beq
|4'\rangle=\cos(\theta)|{\bf n}_{3}\rangle\otimes |e_{3}\rangle
-\sin(\theta)\exp(i\beta)|{\bf n}_{4}\rangle\otimes |e_{4}\rangle.
\eeq
As such, via the creation of an arbitrary superposition of two degenerate target states,
our control scheme also offers a means for the creation of arbitrary quantum
entanglement between the rovibrational and electronic degrees of freedom.  The arbitrary quantum entanglement in molecular systems
thus created may be of importance for quantum information science \cite{qcbook}.
In addition, if the created superposition state $|4'\rangle$ can be used as an initial 
state in
other processes, such as for molecular scattering, then the two participating degenerate states $|3\rangle$ 
and $|4\rangle$ will be able to
interfere with each other. The associated quantum interference effects can be 
manipulated by changing the parameters
$\theta$ and $\beta$, giving rise to an opportunity for
quantum interference control
of collision events \cite{shapiroprl,gongjcp}.

\section{Numerical Examples}

\begin{figure}[ht]
\begin{center}
\epsfig{file=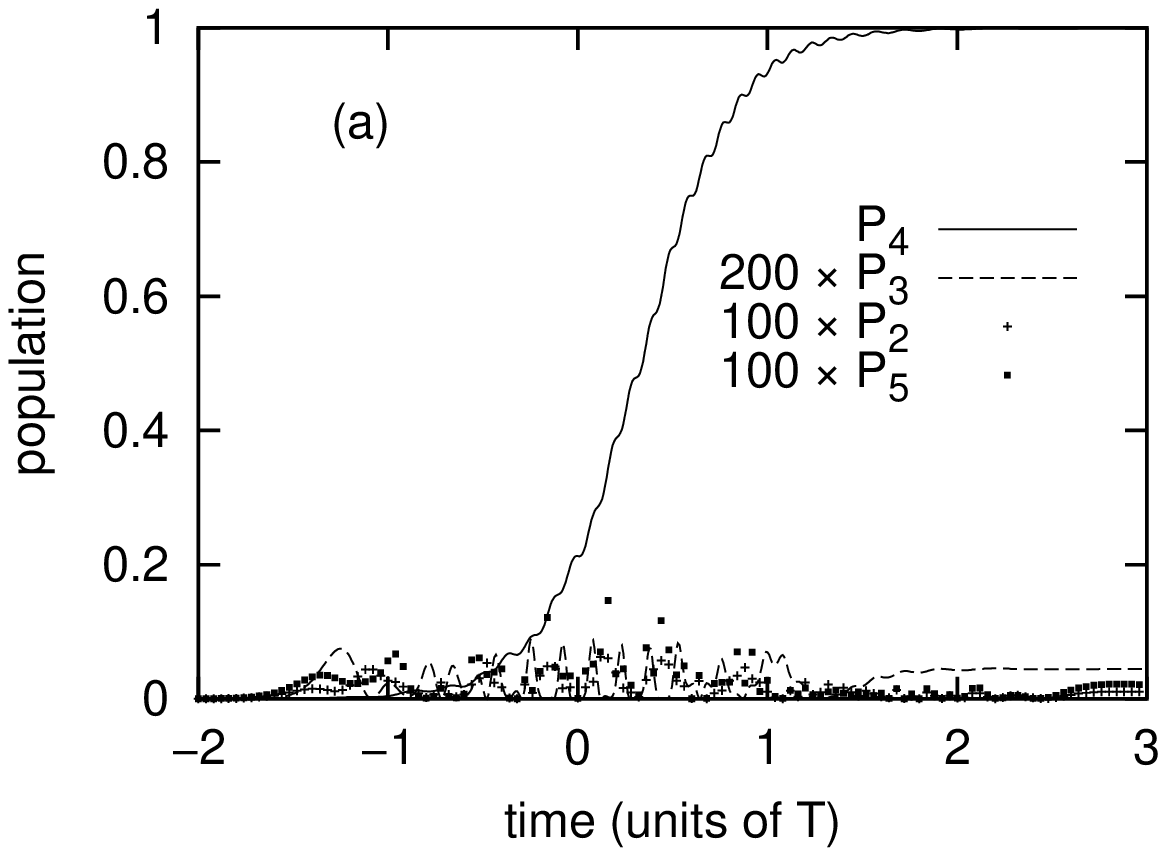,width=5.5cm, angle=270}
\epsfig{file=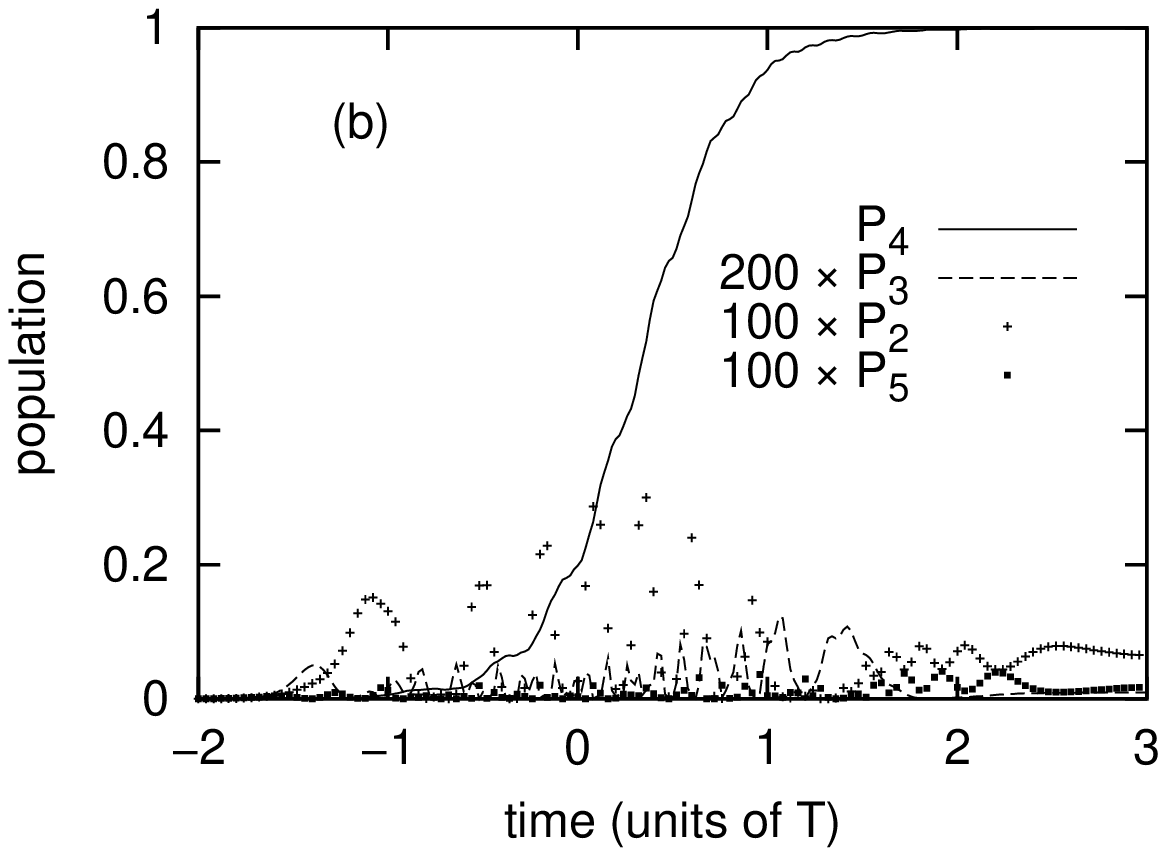,width=5.5cm,angle=270}
\end{center}
\caption{ 
Numerical examples of the dynamics of adiabatic population transfer in the five-level four-pulse phase-sensitive extended STIRAP scheme.
Shown here is the time dependence of the population on state $|k\rangle$ (denoted  $P_{k}$).
$\Omega_{P}T=40 f_{P}(t)$, $\Omega_{S4}T=(20+20i) f_{S}(t)$,  $\Omega_{B4}T=(30+20i) f_{B}(t)$,
$\Omega_{C}T=(10+50i)f_{C}(t)$, with $f_{P}(t)=f_{C}(t)=\exp[-(t-T)^2/T^2]$ and $f_{S}(t)=f_{B}(t)=\exp(-t^2/T^2)$.
In case (a) $\Omega_{S3}T=30 f_{S}(t)$ and $\Omega_{B3}T=20 f_{B}(t)$; in case (b)
$\Omega_{S3}T=10 f_{S}(t)$ and $\Omega_{B3}T=(80i)f_{B}(t)$. Note that during the population transfer
$P_{2}$, $P_{3}$, and $P_{5}$ are negligible at all times.}
\label{fig2}
\end{figure}

In this section we provide some numerical examples of the control of population transfer in our five-level
four-pulse scheme. Consider first the results shown in Figs. 2 and 3 where the control goal is to completely
suppress the yield of state  $|3\rangle$. To obtain 
the results shown in Fig. 2,  we assumed that the product states do not decay and that
\beq
\Omega_{P}T&=&40f_{P}(t), \\
\Omega_{S4}T&=&(20+20i)f_{S}(t),\\
\Omega_{B4}T&=&(30+20i)f_{B}(t),\\
\Omega_{C}T&=&(10+50i)f_{C}(t),
\eeq
with 
\beq
f_{P}(t)&=&f_{C}(t)=\exp[-(t-T)^2/T^2], \label{fp}\\
f_{S}(t)&=&f_{B}(t)=\exp(-t^2/T^2). \label{fs}
\eeq
With these choices it is seen that condition (\ref{con1}) is met,
so the yield of state $|4\rangle$ is expected to be almost 100\% and the yield of state $|3\rangle$
is expected to be almost zero, irrespective of the properties of state $|3\rangle$.  Two cases with drastically different 
transition dipole moments that are related to state $|3\rangle$ are considered.
In case (a) $\Omega_{S3}T=30 f_{S}(t)$ and $\Omega_{B3}T=20 f_{B}(t)$, and in case (b)
$\Omega_{S3}T=10 f_{S}(t)$ and $\Omega_{B3}T=(80i)f_{B}(t)$.  As seen from the detailed time dependence of the population (denoted $P_{k}$)
on state $|k\rangle$ $(k=2-5)$ shown in Fig. 2,
the results in both cases (a) and (b) confirm our previous analysis. In particular, 
the population transfer to state $|4\rangle$ is complete,
the maximal $P_{3}$ during the population transfer
is less than 0.05\% in case (a) and less than 0.07\% in case (b), and in both cases
$P_{2}$ and $P_{5}$ are less than 0.4\% at all times.
Of course, due to some weak nonadiabatic effects associated with the dynamics of population transfer, 
$P_{2}$, $P_{3}$, and $P_5$ are not perfectly zero.
As in the three-level version of STIRAP, using stronger laser fields or larger pulse durations will greatly enhance adiabaticity and
therefore further suppress $P_{2}$, $P_{3}$, and $P_{5}$. For example, we have checked that if all the Rabi frequencies or the pulse widths are
increased by a factor of ten, then the maximal $P_{2}$, $P_{3}$, and $P_{5}$ during the population transfer will be further decreased by
more than one order of magnitude.  In addition, for given field intensity and pulse duration
one may optimize the time delay between the laser pulses to obtain the smallest nonadiabaticity.

\begin{figure}[ht]
\begin{center}
\epsfig{file=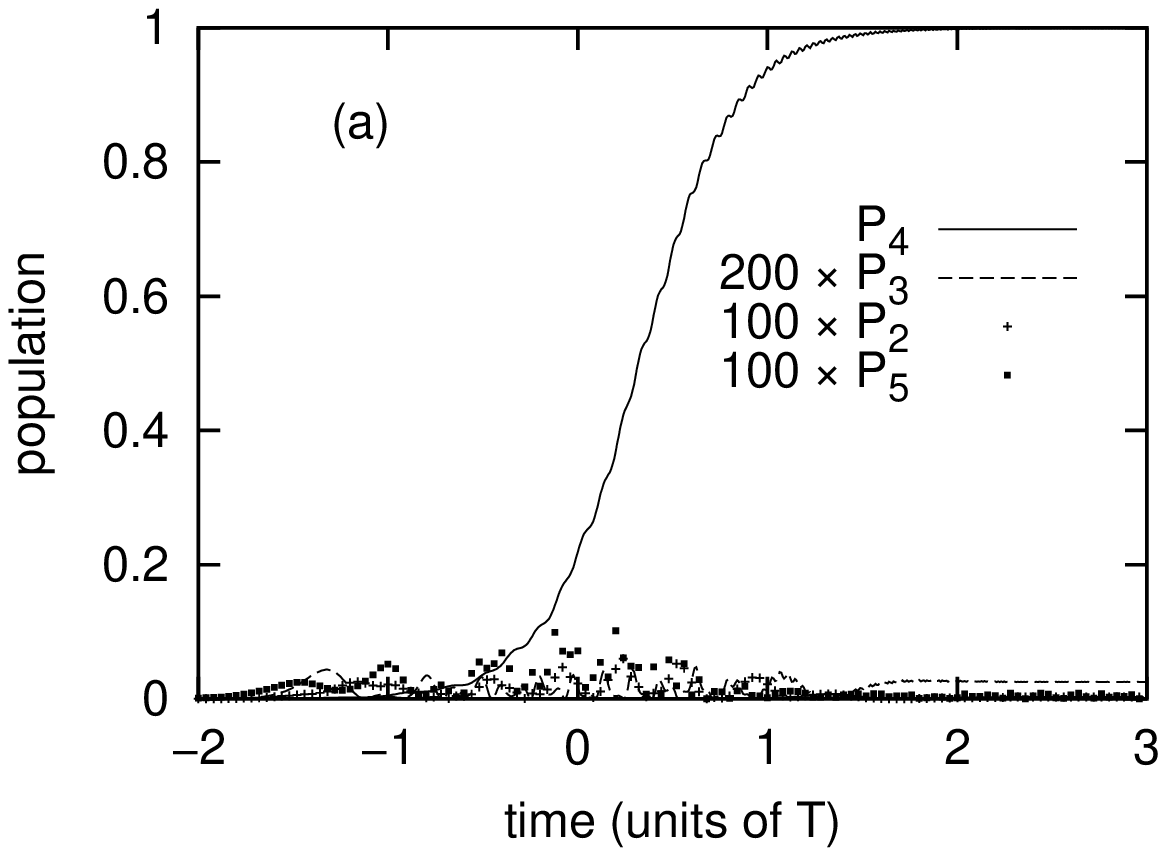,width=5.5cm, angle=270}
\epsfig{file=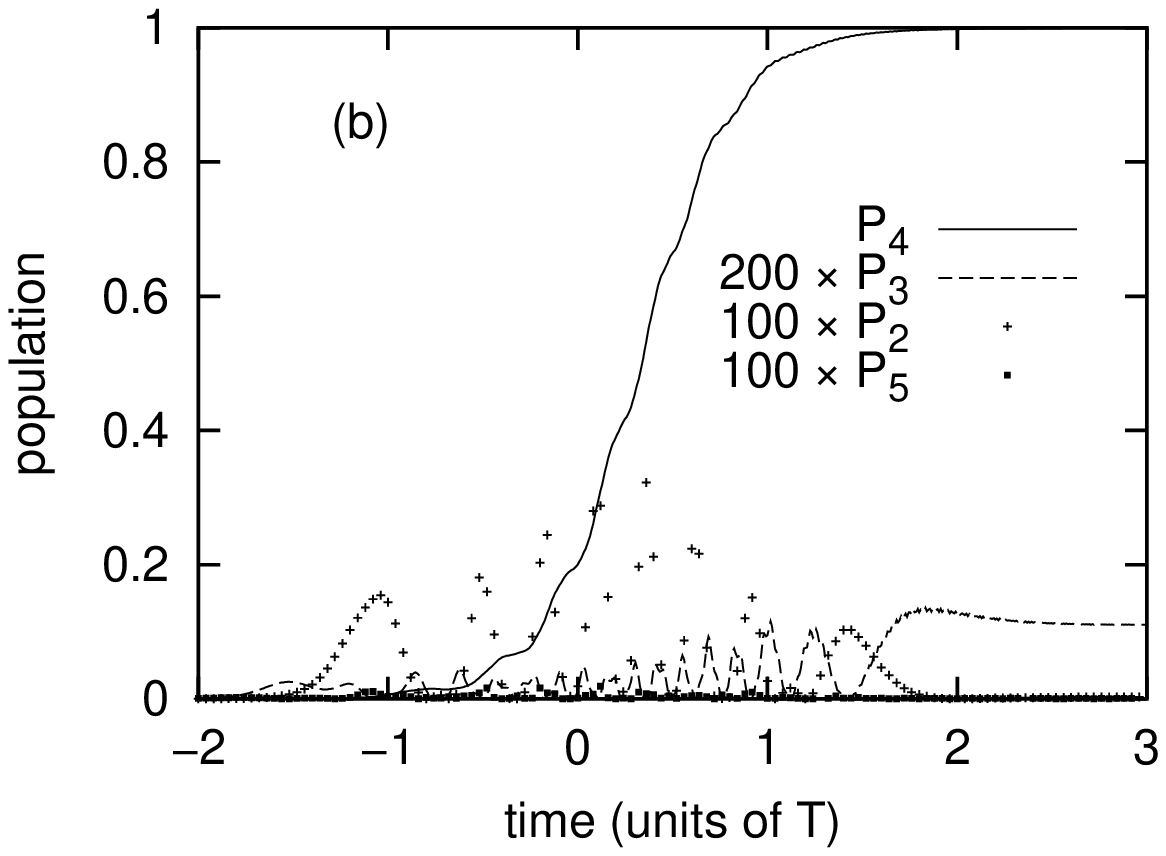,width=5.5cm,angle=270}
\end{center}
\caption{Same as in Fig. 2 except that $\Omega_{C}T$ and the pulse envelope functions are different.
Here $\Omega_{C}T=(10+50i)f_{C}(t)\exp(2.0)$,
$f_{P}(t)=\exp[-(t-T)^2/T^2]$,  $f_{S}(t)=\exp(-t^2/T^2)$, $f_{B}(t)=\exp[-0.5(t-0.5)^2/T^2]$,
and $f_{C}(t)=\exp[-0.5(t-2.5T)^2/T^2]$.  As in Fig. 2, cases (a) and (b) assume different
properties of state $|3\rangle$.
}
\label{fig3}
\end{figure}

Figure 3 displays the analogous results for a different choice of the pulse envelope functions. Here the branching pulse is in the middle
of the pump and the Stokes pulses, and the control pulse is the most delayed pulse.  As discussed above, condition (\ref{con1}) can also be easily met.
Except for $\Omega_{C}T=(10+50i)f_{C}(t)\exp(2.0)$, other 
Rabi frequencies here are chosen to have the same peak values as those used in Fig. 2, with their envelope functions given by
\beq
f_{P}(t)&=&\exp[-(t-T)^2/T^2],  \\
f_{S}(t)&=&\exp(-t^2/T^2), \\
f_{B}(t)&=&\exp[-0.5(t-0.5T)^2/T^2],\\
f_{C}(t)&=&\exp[-0.5(t-2.5T)^2/T^2]. 
\eeq
As seen in Fig. 3, the control obtained is almost perfect, for both cases (a) and (b)
that assume different properties of state $|3\rangle$.

\begin{figure}[ht]
\begin{center}
\epsfig{file=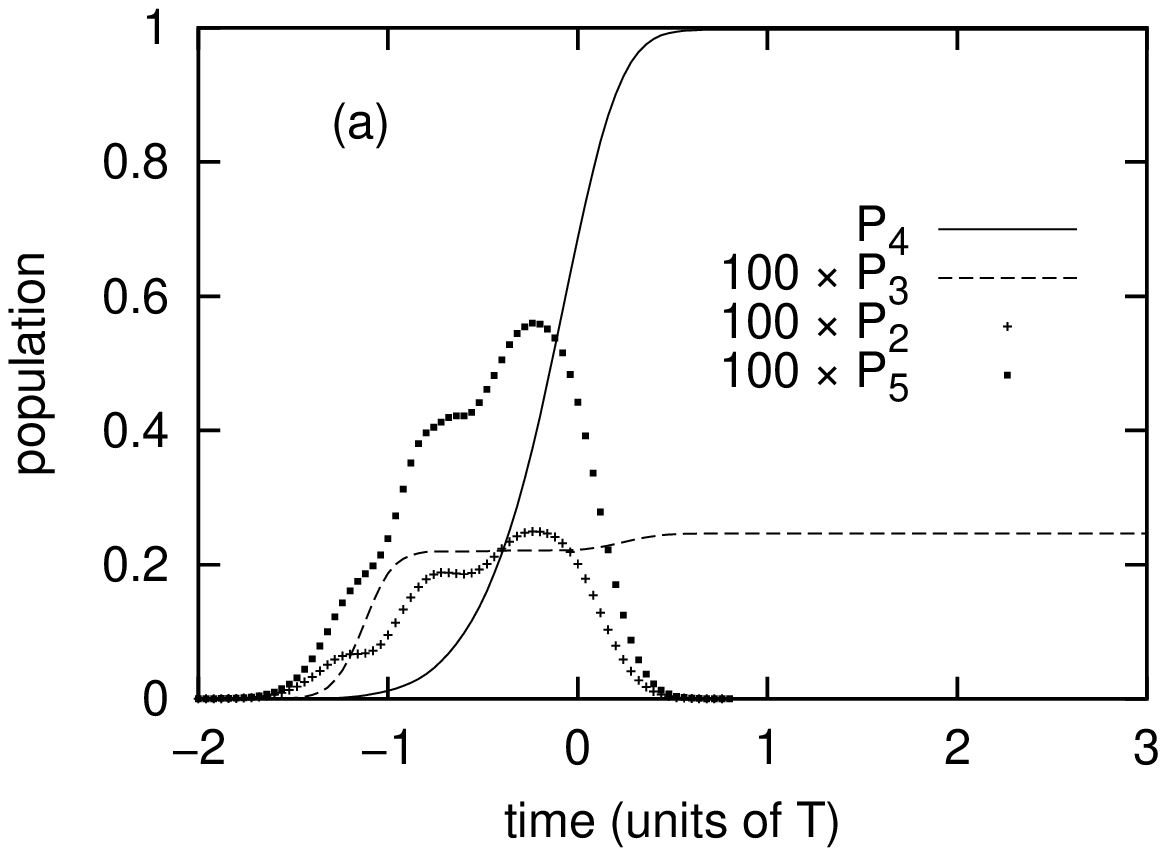,width=5.5cm, angle=270}
\epsfig{file=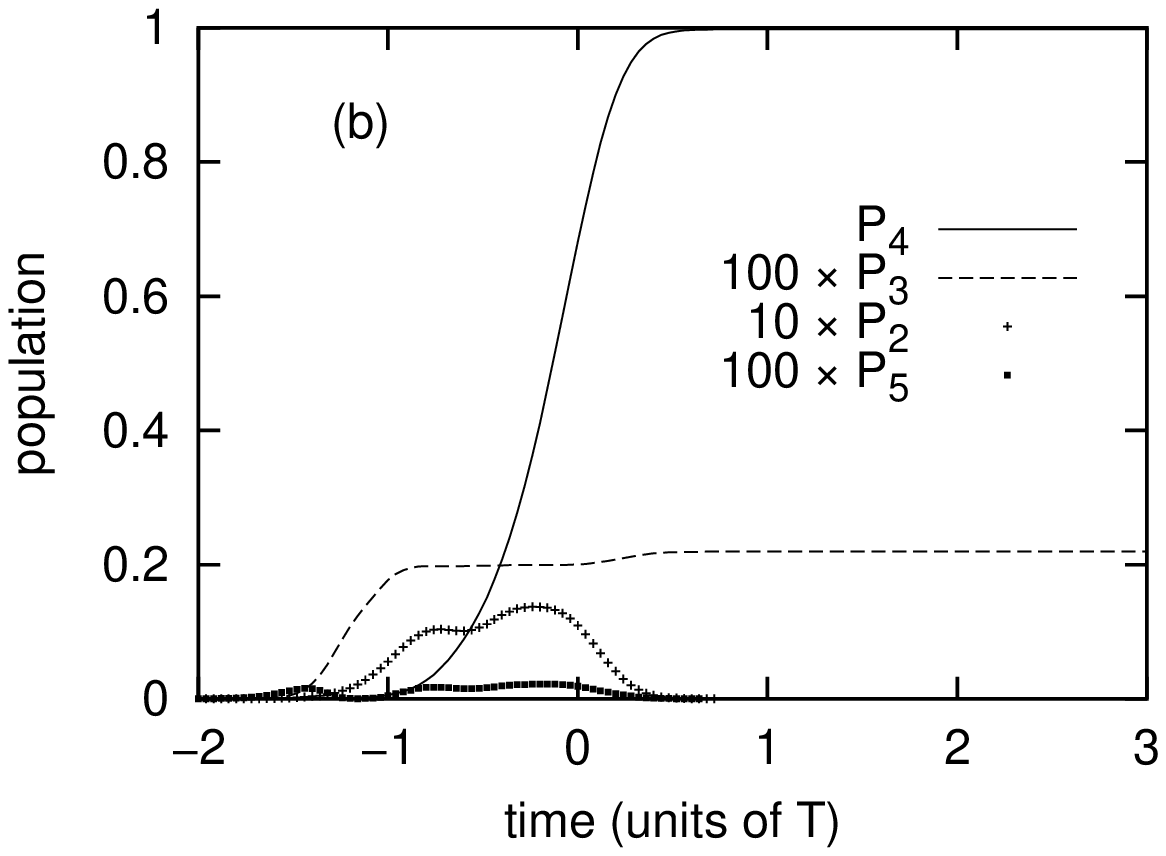,width=5.5cm,angle=270}
\end{center}
\caption{Same as in Fig. 2 except that the degenerate product states $|3\rangle$ and $|4\rangle$ have a common finite lifetime that is ten times smaller
than the pulse duration.}
\label{fig4}
\end{figure}

As mentioned above, the almost complete suppression of $P_{3}$ at all times suggests that the control of population transfer
can hold when the product states rapidly decay. 
This is also confirmed numerically. For example, in Fig. 4 we show the population dynamics when both the product states have
a lifetime
that is one order of magnitude smaller than the pulse duration, with all other parameters same as in Fig. 2.
It is seen that
although the final value of $P_{3}$ is larger than in the non-decaying case in Fig. 2, it is still less than $0.3\%$.  
It is also seen that there is an increase in  $P_{2}$ and $P_{5}$ at intermediate times (but they are still negligible).
This result  is understandable because in the presence of decaying product states
the dynamics is not fully coherent.  The situation can always be further improved by use of stronger laser fields. 

\begin{figure}[ht]
\begin{center}
\epsfig{file=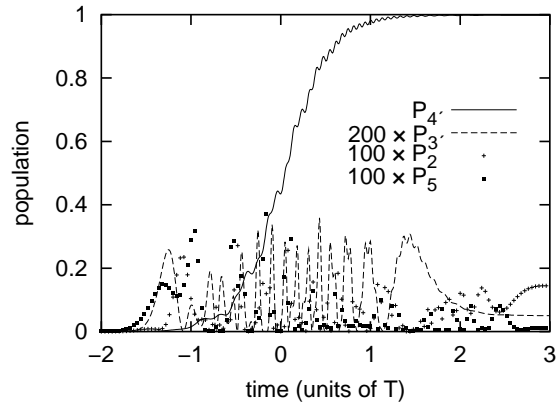,width=5.5cm, angle=270}   
\end{center}
\caption{Dynamics of adiabatic population tranfer from the initial state $|1\rangle$ to a superposition of two degenerate target states given by
$|4'\rangle=(|3\rangle + i|4\rangle)/\sqrt{2}$. $P_{4'}$ and $P_{3'}$ represent the population on states $|4'\rangle$ and
$|3'\rangle=(|3\rangle - i|4\rangle)/\sqrt{2}$, $P_{2}$ and $P_{5}$ represent the population on states $|2\rangle$ and $|5\rangle$.
The pulse envelope functions are given by Eqs. (\ref{fp}) and (\ref{fs}),  $\Omega_{C}T=(80+40i)f_{C}(t)$, and  all  other Rabi frequencies
are chosen to be the same as those used in case (a) in Fig. 2.
}
\label{fig5}
\end{figure}

We have also numerically examined our approach to the creation of an arbitrary superposition of states $|3\rangle$ and $|4\rangle$ whose lifetimes are 
much longer than the pulse duration.
One computational
example is shown in Fig. 5, where the target state is  $|4'\rangle=(|3\rangle + i|4\rangle)/\sqrt{2}$. Then the unwanted superposition state
is $|3'\rangle=(|3\rangle - i|4\rangle)/\sqrt{2}$.
The pulse envelope functions are given by Eqs. (\ref{fp}) and (\ref{fs}),  and except for $\Omega_{C}$, all  other Rabi frequencies
are chosen to be the same as those used in case (a) in Fig. 2.  Given these conditions, we find that in order to induce a node
on state $|3'\rangle$ the control pulse should be designed to give $\Omega_{C}T=(80+40i)f_{C}(t)$.   As shown in Fig. 5, this indeed gives almost 100\% yield of the superposition state
$|4'\rangle$, with the population on states $|2\rangle$, $|5\rangle$, and $|3'\rangle$ being negligible at all times.
Similar results are obtained for other superposition states as target states.

\section{Concluding remarks}

 Using a five-level model, we have shown that complete control of the population transfer branching ratio between two degenerate product states
is achievable using four laser pulses with carefully designed relative phases and envelope functions.
The complete control of population transfer stems from a 
peculiar light-matter interaction in the presence of strong laser fields:
the control mechanism lies in a three-node null eigenstate that can correlate with
an arbitrary superposition of two degenerate target states.
Potential applications of our results include absolute
selectivity of product formation in a photochemical reaction
and the creation of an arbitrary superposition of two degenerate target states.

The most fascinating and somewhat surprising  feature of our five-level four-pulse scheme is that
it allows for the prediction of complete suppression of the yield of one unwanted state of two degenerate product states without
studying its properties beforehand.  This feature of the scheme 
likely has some deep implications.  On the other hand, it becomes very interesting
to ask if there exists a control scheme that can guarantee
100\% yield of one of two degenerate product states with its main
properties (such as its transition dipole moments) unknown to us. 
We are not yet ready to give a conjecture here, but believe 
that if the answer is no then the reason could be related to
the quantum no-cloning theorem \cite{zurek}.

This study also suggests that  STIRAP-like dynamics can shed considerable light on, and may provide a systematic solution to, the
formal question of controllability in a multi-level quantum system without or with constraints.  In particular,
the availability of
a complete population transfer pathway that never populates a certain number of states can be examined
by studying the existence of a single multi-node dressed
eigenstate of the system
in the presence of control fields.  If, by designing the control fields,
such a multi-node dressed eigenstate can be made to exist and can correctly
correlate with the initial and target states,
then the controlled dynamics in the adiabatic limit gives a specific
physical solution to the generation of complete population transfer.

Previous quantum control approaches based on STIRAP-like dynamics are independent of laser phases.  As shown in this work
and in Ref. \cite{kis},
this is not always the case.  Rather, one may take advantage of the laser-phase-dependent quantum
interference
between different STIRAP-like sub-processes to achieve more dramatic goals in the laser control of atomic and molecular processes.  
It is our hope that in the near future
extensions of our five-level four-pulse
extended STIRAP scheme can provide us other amazing opportunities for
quantum control.

This work was supported by the National Science Foundation.

\end{document}